\documentclass[twocolumn,prl,showpacs,multicol,amsmath,amssymb]{revtex4-2}
\usepackage{amssymb}
\usepackage{amsmath}
\usepackage{graphicx}
\usepackage{epstopdf}
\usepackage{bm}
\usepackage{gensymb}

\newcommand\redout{\bgroup\markoverwith{\textcolor{red}{\rule[.5ex]{2pt}{0.4pt}}}\ULon}
\usepackage[colorlinks=true,citecolor=blue]{hyperref}
\hypersetup{colorlinks=true,citecolor=blue,linkcolor=red,urlcolor=blue}

\graphicspath{{.}{./plot/}}

\begin{document}

\title{Superconductor Qubits Hamiltonian Approximations Effect on Quantum State Evolution and Control}

\author{Javad Sharifi}\email{sharifi@qut.ac.ir, jv.sharifi@gmail.com}
\affiliation{Qom University of Technology, Qom 37181-46645, Iran}

\date{\today}


\begin{abstract}
Quantum state on Bloch sphere for superconducting charge qubit, phase qubit and flux qubit for all time in absence of external drive is stable to initial state.  By driving the qubits, approximation of charge and flux Hamiltonian lead to  quantum state rotation in Bloch sphere around an axis completely differ  from rotation vector of exact Hamiltonian. The trajectory of quantum state for phase qubit for  approximated and exact Hamiltonian is the same but the expectation of quantum observable has considerable errors as two other qubits. microwave drive control is designed for approximated Hamiltonian and  exerted on actual systems and shows completely different trajectory with respect to desired trajectory. Finally a nonlinear control  with external $\mu$V voltage control and nA current control  is designed for general qubit which completely stabilizes quantum state toward a desired state. 
\end{abstract}

\maketitle

\section{Introduction}
 Quantum engineering harness the quantum state, estimate and eliminate environmental noise on open quantum systems, correction of quantum errors, all for quantum computing. Excellent quantum engineering framework researches are carried for quantum computing applications with trapped ions\cite{cirac1995quantum,blatt2008entangled}, spin qubit control\cite{pla2013high,sharifi2011lyapunov,pla2012single}, quantum optical control of semiconductor quantum dots\cite{bonadeo1998coherent,chen2004theory,greilich2006optical}, superconducting qubits\cite{wallraff2005approaching,campagne2013persistent,leonard2019digital,wallraff2004strong}, etc. Among different approach to quantum computers, since superconductor circuits can be fabricated and control based on current technologies has attracted attention of both researches\cite{houck2012chip,fedorov2012implementation,mariantoni2011implementing, ofek2016extending, campagne2018deterministic} and IC makers, IT companies such as IBM, Intel,Google, Microsoft, etc to make quantum processor with exceptional computational performance with respect to conventional processors. Achieving to this goal, the quantum sate of  each qubit as fundamental building block of quantum computation must be controlled precisely toward desired state and remain stable at that state. 

For simulations, the QuTip, Numpy, Symptom,Matplotlib python packages are applied and all of data are from a table at \cite{wendin2017quantum}. At first, we  introduce the Hamiltonian of basic superconductor qubits and their approximations, then by simulations show that two systems, rotate along different trajectories on Bloch sphere and with different quantum observable expectations, then we endevour to design microwave controller for approximated system and show that this control leads to wrong results with respect to desired trajectories. Finally we design Lyapunov control for general charge-phase-flux nonlinear circuit which converge  to desired state. 

 By the Schrödinger equation  $i\hslash\frac{d|\psi\rangle}{dt}=H|\psi\rangle$ and its unitary evolution $U_{t}=exp(\frac{{-i}}{\hslash}\int_0^t H(t)dt)$ then the solution is $|\psi_{t}\rangle=e^{\frac{-i}{\hslash}Ht}|\psi_{0}\rangle=U_{t}|\psi_{0}\rangle$ ,let X be a hermitian observable, the evolution of observable at time t is $X_{t}=U^{\dag}_{t}XU_{t}$ and the mean value of a  observable based on density matrix $\rho_{t}=|\psi_{t}\rangle \langle \psi_{t}|$ is $\langle X \rangle=\mathrm{Tr}(\rho_{t} X)$. The quantum state on Bloch sphere is $|\psi \rangle=\mathrm{cos}(\frac{1}{2}\theta)|0 \rangle+e^{i\phi}\mathrm{sin}(\frac{ 1}{2}\theta)|1 \rangle $ where $ |0 \rangle$ and  $ |1 \rangle$ are qubit eigenbasis or in density matrix form as $\rho_{t}=\frac{1}{2}\begin{pmatrix}
 	1+z&x-iy\\x+iy&1-z
 \end{pmatrix}$, where $(x,y,z)=\vec{r}=(\mathrm{sin}(\theta)\mathrm{cos}(\phi),\mathrm{sin}(\theta)\mathrm{sin}(\phi),\mathrm{cos}(\theta))$ is vector on spherical coordinate, in this form the quantum system is govern by master equation $\dot{\rho}_{t}=\frac{i}{\hslash}[\rho_{t},H]$. The unitary rotation operator for transition of quantum states is $U_{\mathrm{rot}}=\mathbf{R}_{\hat{n}}(\alpha)=e^{-i\frac{\alpha}{2}\hat{n}.\vec{\sigma}}=\mathrm{sin}(\frac{\alpha}{2})-i\mathrm{cos}(\frac{\alpha}{2})\hat{n}.\vec{\sigma}$ with $\vec{\sigma}=\sigma_{x}\hat{x}+\sigma_{y}\hat{y}+\sigma_{z}\hat{z}$ and $\sigma_{x},\sigma_{y},\sigma_{z}$ are Pauli matrices and $\hat{n}=n_{x}\hat{x}+n_{y}\hat{y}+n_{z}\hat{z}$ is the rotation axis and $\alpha=\omega_{q} t$ is rotation angle, $\omega_{q}$ is the angular speed of quantum state vector on Bloch sphere.
 
\section{Superconductor Qubits Evolution}
\begin{figure}[ht]
	\vspace{0.1cm}
	\includegraphics[width=0.65 \linewidth]{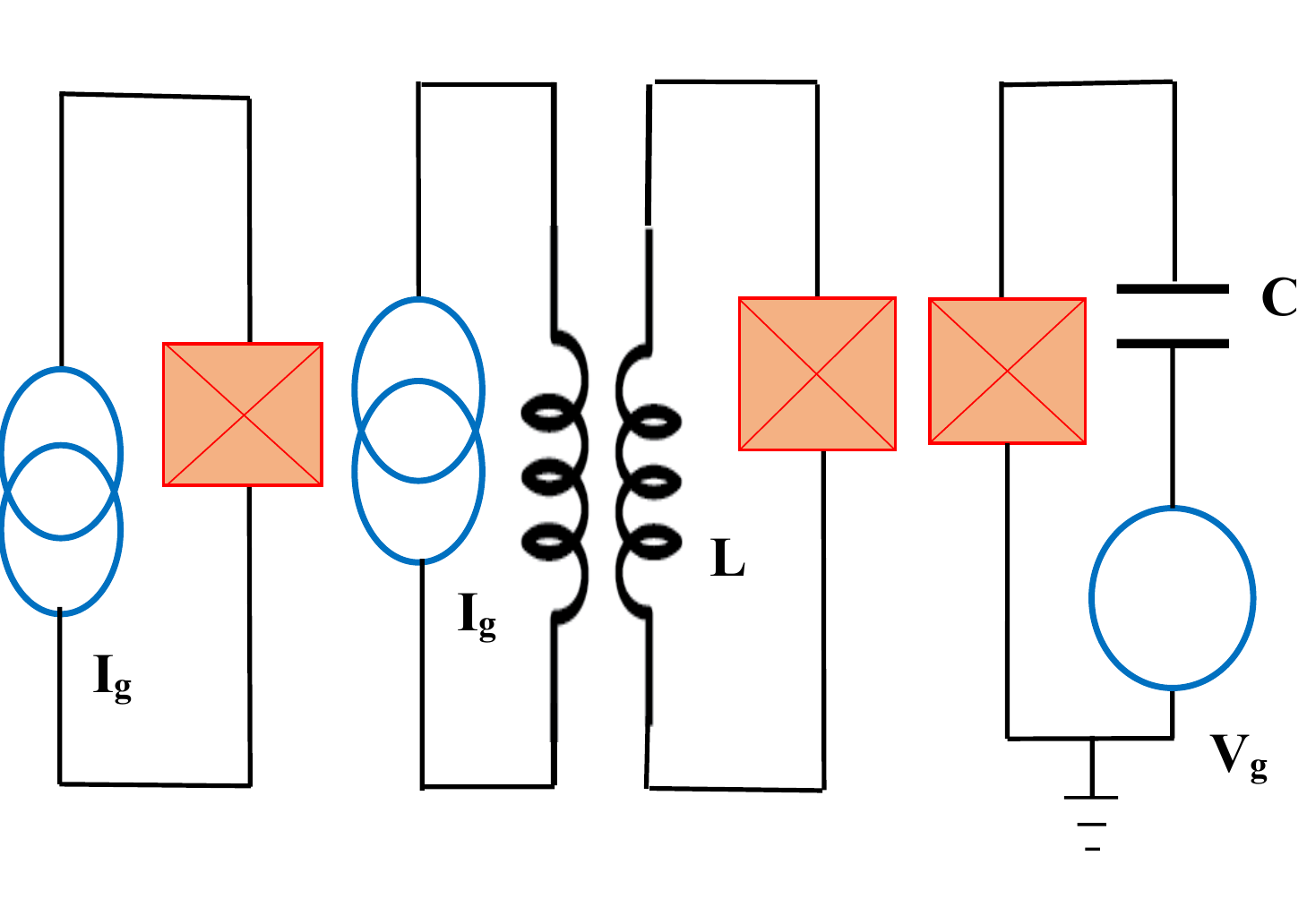}
	\vspace{-0.5cm}
	\caption{Basic superconductor qubits: phase qubit(left), flux qubit(middle) and charge qubit(right) } 
	\label{circuits}
\end{figure}

The fundamental superconductor circuits are depicted in figure\ref{circuits}. 

\begin{figure}[ht]
	\vspace{-0.1cm}
	\includegraphics[width=1 \linewidth]{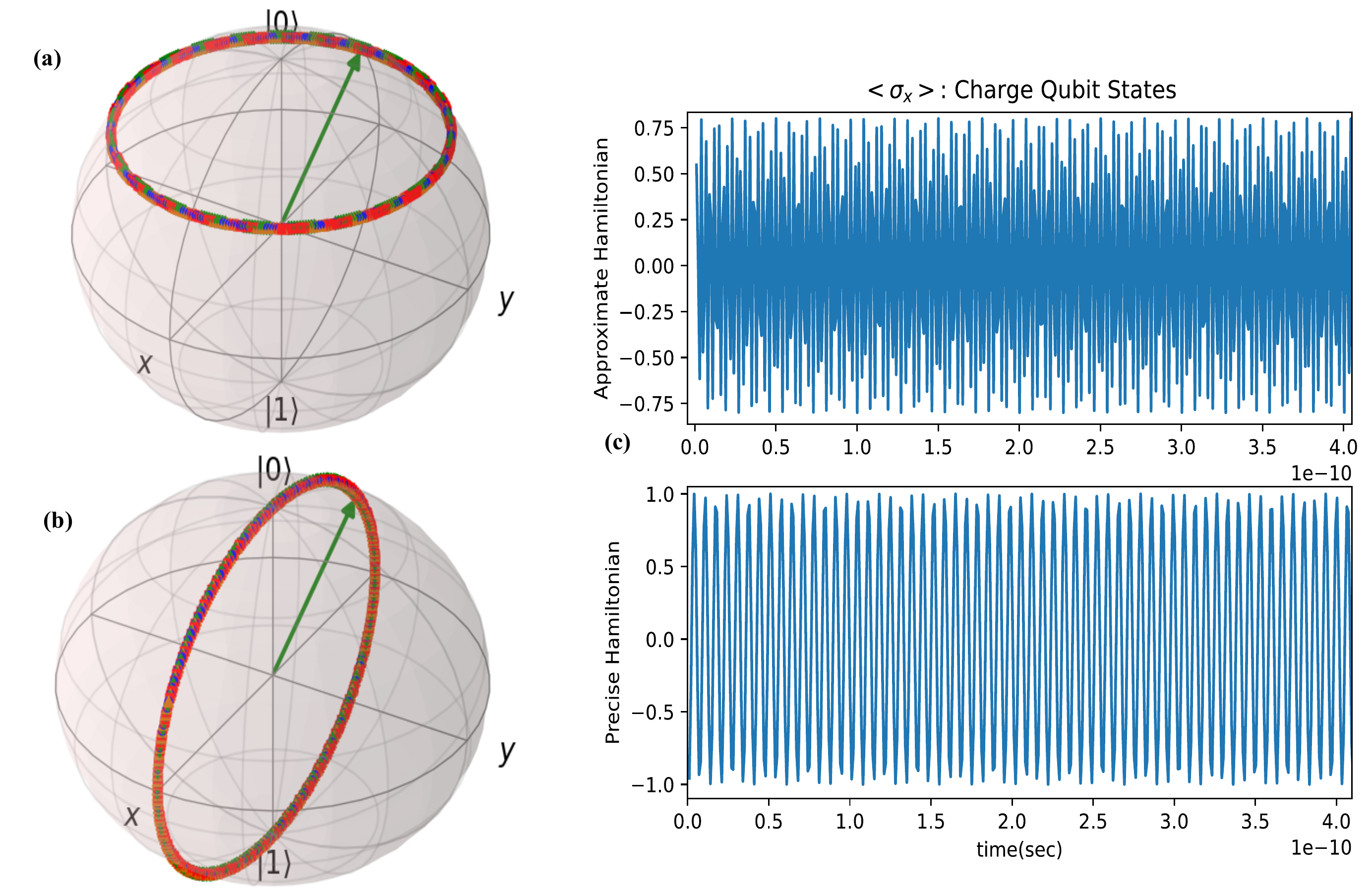}
	\vspace{-0.4cm}
	\caption{Charge qubit state transition without external drive voltage $V_{g}=1$mV, $C_{g}=0.68$fF and $E_{J}=0.018E_{c}$, $E_{c}=7.55\times10^{-23}$J: quantum state on Bloch sphere with approximate Hamiltonian (a) and precise Hamiltonian (b) and the mean value of sates(c)} 
	\label{charge}
\end{figure}

Precise Hamiltonian for charge qubit or C-JJ circuit is \cite{wendin2007quantum}: $\hat{H}_{pre}=E_{c}(\hat{n}-n_{g})^{2}-E_{J}\mathrm{cos}(\hat{\phi})$, in which phase operator $\hat{\phi}$ and charge operator $\hat{n}=-i\hslash\frac{\partial}{\partial{\hat{\phi}}}$ satisfy the commutation relation $[\hat{\phi},\hat{n}]=i\hslash$. $E_{c}=(2e)^2/2(C+C_{g})$ are the charging energy of one 2e Cooper pair which e is electron charge,$E_{J}=\frac{\hslash}{2e}I_{c}$ is Josephson energy, $C_{g},I_{c}$ are Josephson Junction (JJ) capacitor and Josephson critical current, and $n_{g}=\frac{1}{2}C_{g}V_{g}/e$ is the voltage-induced charge on the capacitor C, and it is the control parameter. The approximate hamiltonian with forth order approximation of cosine term and after second quantization is $\hat{H}_{app}=\hbar\omega_{q} a^{\dagger}a+\frac{\beta}{2}a^{\dagger}a^{\dagger}aa$ and resembles a Duffing oscillator. The number operator and phase operator for qubits are set as $\hat{n}=in_{zpf}(a-a^{\dagger})$ and $\hat{\phi}=\phi_{zpf}(a+a^{\dagger})$ where $n_{zpf}=(E_{L_{J0}}/32E_{c})^\frac{1}{4}$ and $\phi_{zpf}=(2E_{c}/E_{L_{J0}})^\frac{1}{4}$ are number and phase zero-point-fluctuation and $E_{L_{J0}}=\hslash^2/(4e^2L_{J0})$. For qubits the creation and annihilation operators are respectively the raising and lowering ladder operators, i.e. $a^{\dagger}=\sigma_{+} , a=\sigma_{-}$. By assumption $E_{c}>>E_{J}$ the Hamiltonian simplifies to\cite{rodrigues2003superconducting}: $\hat{H}_{app}=E_{c}(\frac{1}{2}-n_{g})\sigma_{z}+\frac{1}{2}E_{J}\sigma_{x}$ and then the quantum state evolve by equation $|\psi\rangle=e^{-\frac{it}{\hslash}(E_{c}(\frac{1}{2}-n_{g})\sigma_{z}+\frac{1}{2}E_{J}\sigma_{x})}|\psi_{0}\rangle$ which by comparison to unitary rotation operator, this corresponds to the rotation around an axis on x-z plane on Bloch sphere as is illustrated on figure\ref{charge}-(a). The simulation of state evolution with exact Hamiltonian is shown on figure\ref{charge}-(b), however for zero deriving voltage of charge qubit (JJ-C), the state vector will remain constant to initial state on the surface of Bloch sphere. In figure\ref{charge}-(c), the expectation value of a Pauli matrix , i.e. $\langle \sigma_{x} \rangle$ is plotted. 

\begin{figure}[ht]
	\vspace{-0.1cm}
	\includegraphics[width=1 \linewidth]{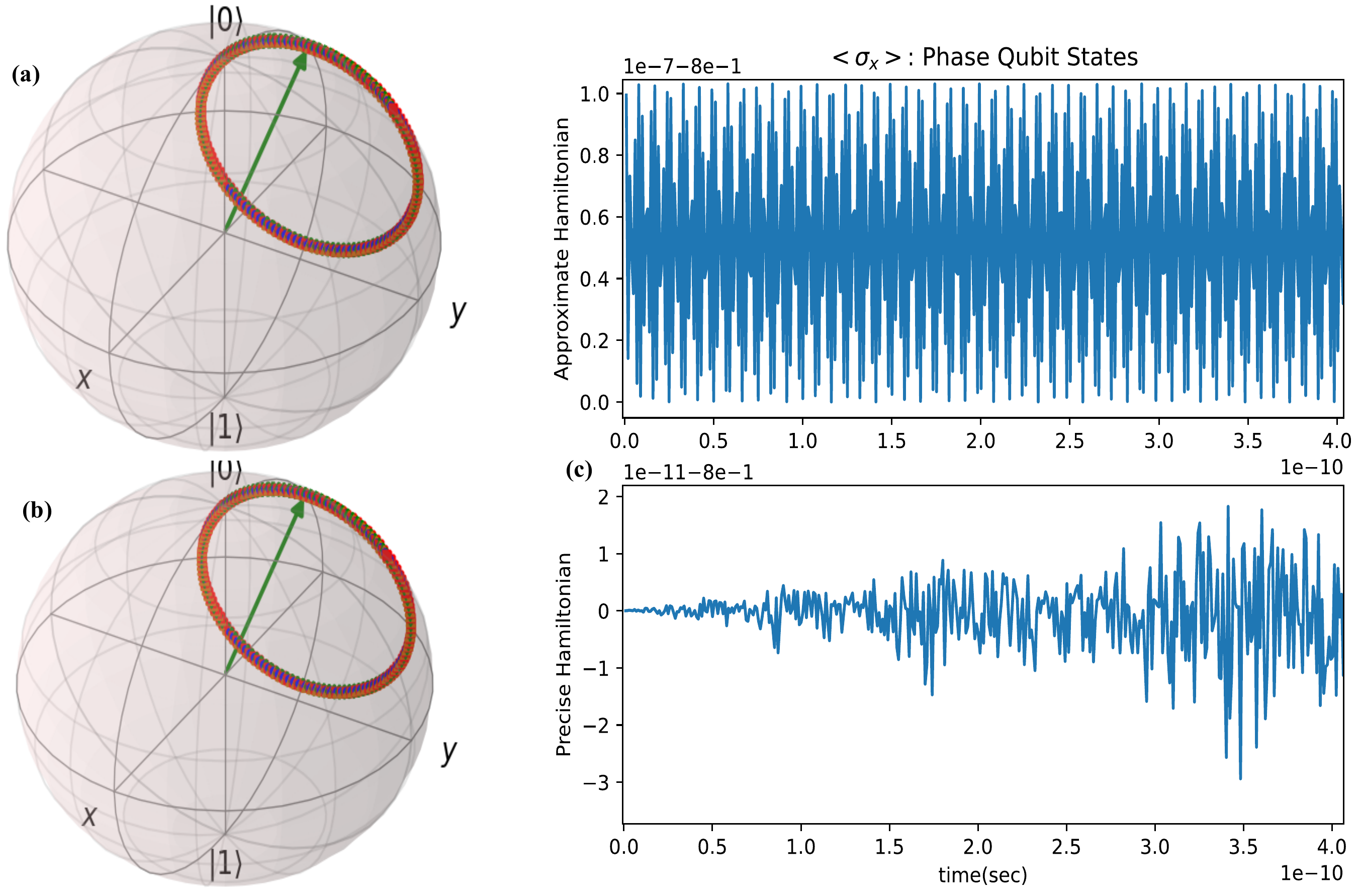}
	\vspace{-0.4cm}
	\caption{Phase qubit state evolutiopn($E_{J}=3.266\times10^{-23}$, $E_{c}=10^{-4}E_{J}$, $\phi_{zpf}=0.0398$, $I_{g}=1$mA): state evolution on Bloch Sphere with approximate Hamiltonian (a) and precise Hamiltonian (b) and the mean value of sates(c)} 
	\label{phase}
\end{figure}

Phase qubit is depicted in right of figure\ref{circuits} and obeys the relation\cite{wendin2007quantum}: $\hat{H}_{pre}=E_{c}\hat{n}^{2}-E_{J}\mathrm{cos}(\hat{\phi})-\frac{\hslash}{2e}I_{g}\hat{\phi}$ where  $I_{g}$ is the control current. The approximate Hamiltonian is $\hat{H}_{app}=-\frac{1}{2}E_{c}\sigma_{z}+(\frac{1}{2}E_{J}-\frac{\hslash}{2e}\phi_{zpf}I_{g})\sigma_{x}$. Simulations of phase qubit evolution both for approximate and precise Hamiltonian is depicted in figure\ref{phase_control}-(a) and (b) respectively that shows the same result trajectory but the expectation value of quantum observable $\langle \sigma_{x}\rangle$ is completely different for two Hamiltonian(figure\ref{phase}-(c)).

\begin{figure}[ht]
\vspace{-0.1cm}
\includegraphics[width=1 \linewidth]{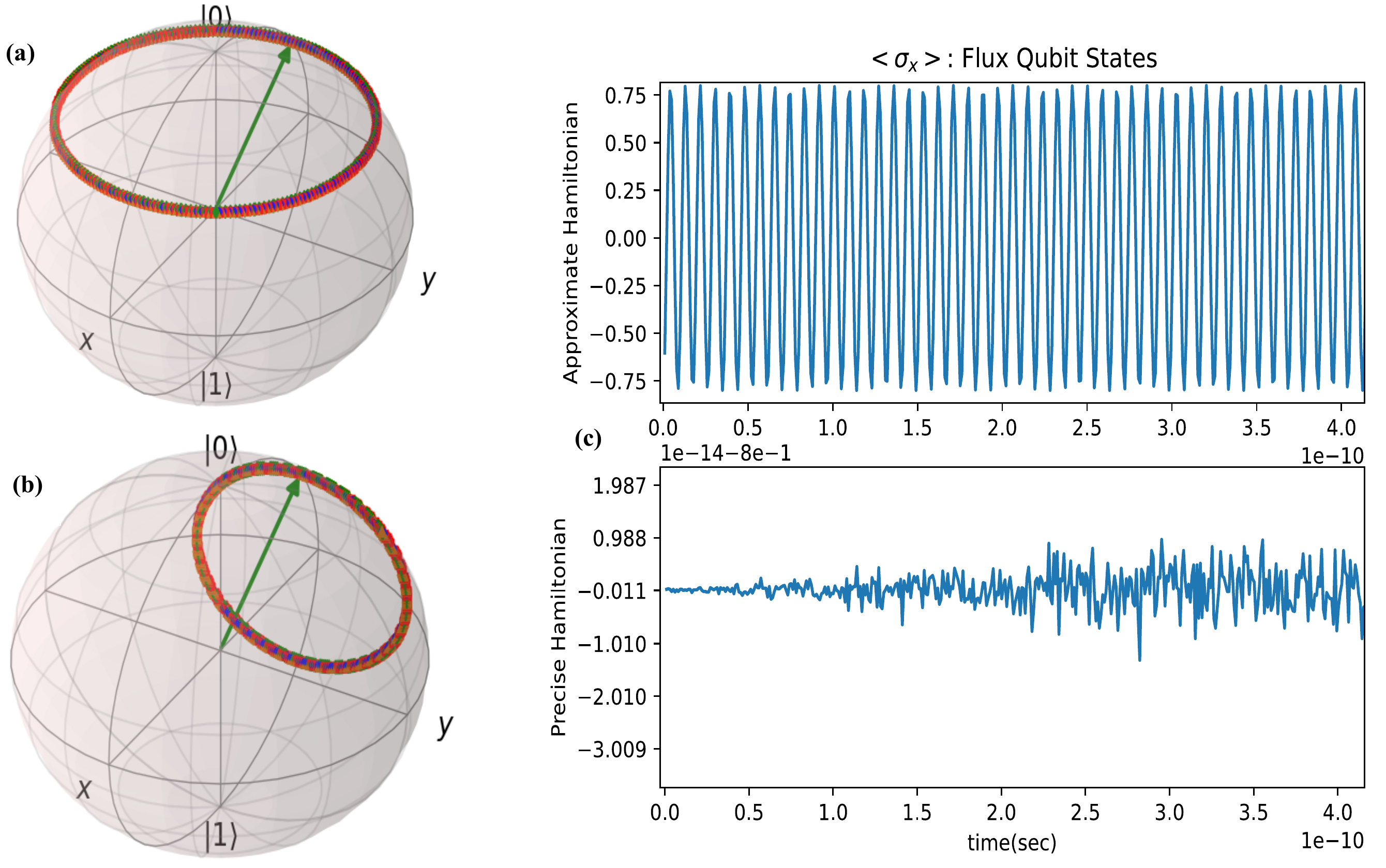}
\vspace{-0.4cm}
\caption{Flux qubitsState transition: State Transition on Bloch Sphere with Approximate Hamiltonian (Top) and Precise Hamiltonian (Bottom) and the mean value of sates(c): $E_{J}=6.017\times10^{-23}$J , $E_{c}=1.711\times10^{-23}$J} 
\label{flux}
\end{figure}

Flux qubit (L-JJ) is depicted in middle of figure\ref{circuits} and obeys the relation\cite{wendin2007quantum}: $\hat{H}_{pre}=E_{c}\hat{n}^{2}-E_{J}\mathrm{cos}(\hat{\phi})+\frac{1}{2}E_{L} (\hat{\phi}-\phi_{e})^2$ that $E_{L}=\hslash^2/(4e^2L)$ and $\phi_{e}$ is the control parameter.  Approximate  Hamiltonian is $\hat{H}_{app}=-\frac{1}{2}E_{c}\sigma_{z}+(\frac{1}{2}E_{J}-E_{L}\phi_{zpf}\phi_{e})\sigma_{x}$. For approximate Hamiltonian, the trajectory of quantum state on Bloch sphere rotate around z-axis (figure\ref{flux}-(a)) but trajectory rotation change for precise model of Hamiltonian (figure\ref{flux}-(b)) and also the expectation $\langle \sigma_{x} \rangle$ of both Hamiltonian is different (figure\ref{flux}-(c)). As we found from this simulations is that approximating the superconducting qubits lead to wrong  state evolution and considerable error in expectation of quantum observable.

\section{Effect on Qubits Control}
By the equation of Hamiltonian in previous section it is figured out that by constant control signal, it is not possible to control quantum state vector over whole space through rotation of quantum state on three direction, for example by approximate Hamiltonian of charge qubit we can control the rotation of state around z-axis. one approach to overcome this shortage is by microwave driving oscillator (Rabi driving)\cite{saito2006parametric, kutsuzawa2005coherent,naaman2017josephson,krantz2019quantum} which practically produces by arbitrary wave generator (AWG). Here, we drive the charge qubit by the signal which contains both Rabi drive and non-oscillating dc signal to fully control the quantum state on three rotation directions, assume the signal as $V_{g}(t)=V_{ac}+V_{dc}=V_{0}s(t)\mathrm{sin}(\omega_{c}t+\lambda)+V_{g0}$ which $\omega_{c}$ is control signal frequency, the initial non-rotating frame control hamiltonian is $\hat{H}_{c}=-E_{c}n_{g}\sigma_{z}$ then the Hamiltonian of rotating part is given by $\hat{H}_{rf}=U^{\dag}_{0}kV_{0}s(t)\mathrm{sin}(\omega_{c}t+\lambda)\sigma_{z}U_{0}$ where $U_{0}=e^{-\frac{it}{\hslash}(\frac{1}{2}E_{c}\sigma_{z}+\frac{1}{2}E_{J}\sigma_{x})}=e^{-\frac{i}{2}\omega_{z}t\sigma_{z}-\frac{i}{2}\omega_{x}t\sigma_{x}}$ and $k=(-\frac{1}{2e}C_{g}E_{c})$, multiplication in $\hat{H}_{rf}$ has sine and cosine terms with frequencies $\omega_{c}-(\omega_{z}-\omega_{x})$ , $\omega_{c}+\omega_{z}+\omega_{x}$ , $\omega_{c}+\omega_{z}-\omega_{x}$ , $\omega_{c}-\omega_{z}-\omega_{x}$ and since for charge qubit $\omega_{z}>>\omega_{x}$ by rotating wave approximation, only first term is hold and other three fast rotating terms and constant offset can be ignored, then the rf-controlled Hamiltonian is:
\begin{align}
\begin{aligned}
\hat{H}_{rf}=\frac{_1}{^{8}}kV_{0}s(t)(\mathrm{sin}(\delta\omega t+\lambda)\sigma_{x}-2\mathrm{cos}(\delta\omega t+\lambda)\sigma_{y}\\
+\mathrm{sin}(\delta\omega t+\lambda)\sigma_{z})\label{equ:1}
\end{aligned}
\end{align}
which in this equation $\delta\omega=\omega_{c}-(\omega_{z}-\omega_{x})$, if we set $\delta\omega=0$, only the phase difference between radio frequency with frequency $\omega_{x}$ plays control role and the rf-controlled Hamiltonian simplifies to $\hat{H}_{rf}=\frac{1}{8}kV_{0}s(t)(Q\sigma_{x}-2I\sigma_{y}+Q\sigma_{z})$ which $I=cos(\lambda),Q=sin(\lambda)$, this method is famous method of IQ-mixer and since by this Hamiltonian only it is possible to control quantum state in two rotation because every rotation around x-axis lead to same rotation around z-axis, hence we add constant dc voltage pulse to change the total control propagator operator to:
 \begin{align}
\begin{aligned}
\label{equ:1}
  U_{c}(t)=e^{\frac{-ik}{\hslash}(\frac{V_{0}}{8}\gamma(t)Q\sigma_{x}-\frac{V_{0}}{4}\gamma(t)I\sigma_{y}+(\frac{V_{0}}{8}\gamma(t)Q+V_{g0}t)\sigma_{z})}
\end{aligned}
\end{align}  
$\gamma (t)=\int_0^t s(t)dt$, for simplicity let set $s(t)=1$ and then $\gamma (t)=t $, hence by comparison to rotation propagator it obtains: $\omega_{q}n_{x}=\frac{1}{4\hslash}kV_{0}Q , \omega_{q}n_{y}=-\frac{1}{2\hslash}kV_{0}I , \omega_{q}n_{z}=\frac{1}{4\hslash}kV_{0}Q+\frac{2}{\hslash}kV_{g0}$ then the whole signal parameters are: $\lambda=tan^{-1}(-\frac{2n_{x}}{n_{y}}) , V_{0}=\frac{4\hslash\omega_{q} n_{x}}{kQ} , V_{g0}=\frac{\hslash\omega_{q}}{2k}(n_{z}-n_{x})$. now, consider the control signals aid to transit initial quantum state $|\psi_{0} \rangle$ to final quantum state $|\psi_{f} \rangle$ at finite time $t_{f}$, one approach is to rotate the initial vector to final vector around the around the bisector of two unit vectors $\vec{r_{0}},\vec{r_{f}}$ by angle $\pm\pi$, then:
\begin{align}
\begin{pmatrix}
\hat{n}\\
\alpha
\end{pmatrix}
=
\begin{pmatrix}
\frac{\vec{r_{0}}+\vec{r_{f}}}{|\vec{r_{0}}+\vec{r_{f}}|}\\
\pm \pi
\end{pmatrix}
\end{align}

\begin{figure}[ht]
\vspace{-0.1cm}
\includegraphics[width=0.7 \linewidth]{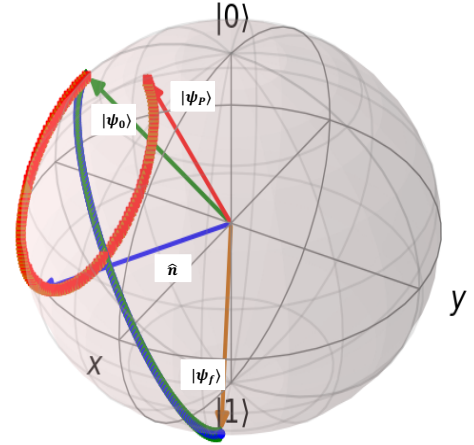}
\vspace{-0.4cm}
\caption{Quantum state control of superconductor charge qubit: $|\psi_{0}\rangle$ is initial state $|\psi_{f}\rangle$ is final desired state. Controlled desired trajectory is depicted in Blue. Exerting the designed controller to main system move trough red trajectory and reach to wrong state $|\psi_{P}\rangle$.} 
\label{charge_control}
\end{figure}

\begin{figure}[ht]
\vspace{-0.1cm}
\includegraphics[width=0.7 \linewidth]{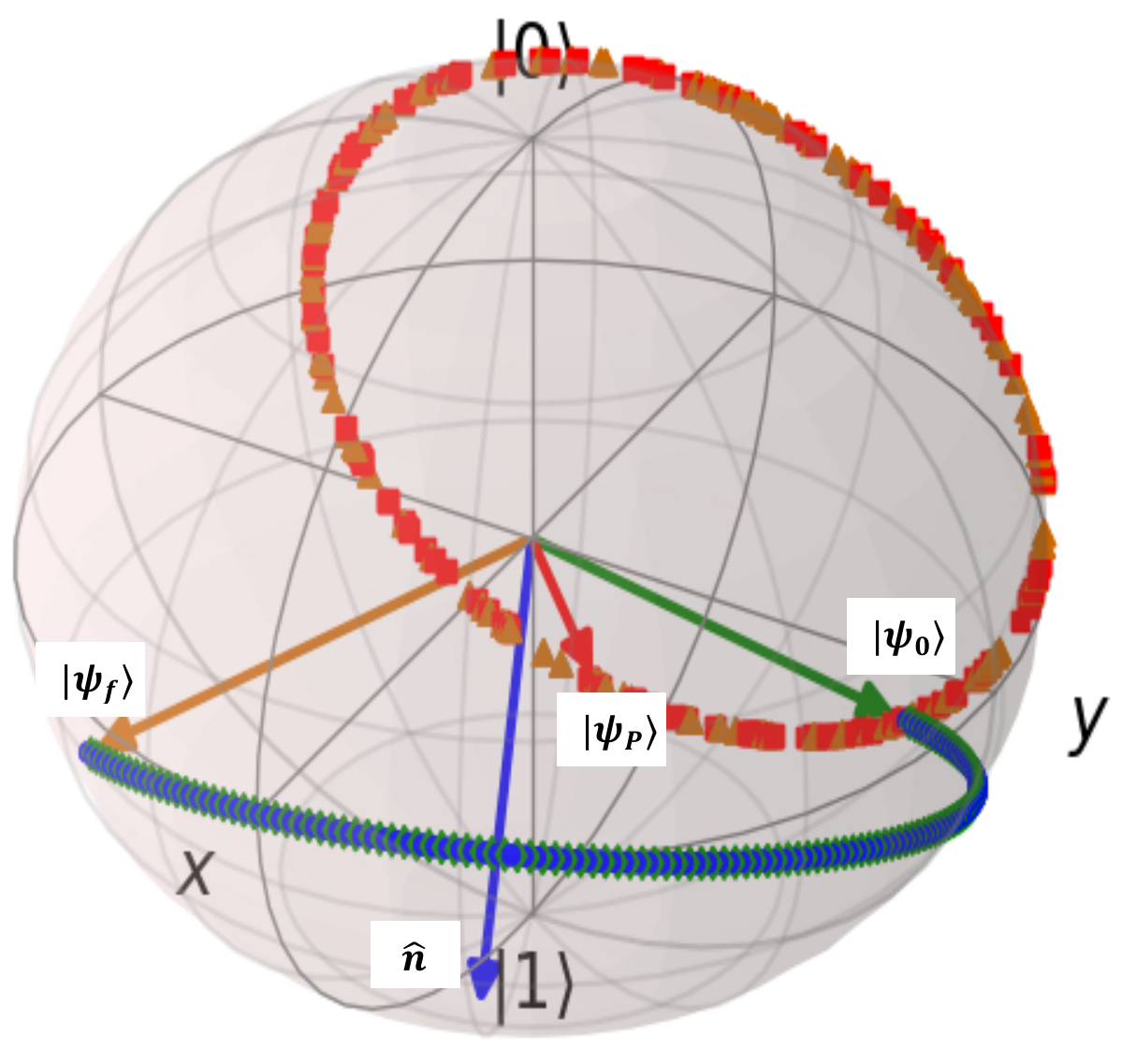}
\vspace{-0.4cm}
\caption{Quantum state control of superconductor phase qubit: $|\psi_{0}\rangle$ is initial state $|\psi_{f}\rangle$ is final desired state. Controlled desired trajectory is depicted in Blue. Exerting the designed controller to main system move trough red trajectory and reach to wrong state $|\psi_{P}\rangle$.} 
\label{phase_control}
\end{figure}

For example, let set the initial and final states as $|\psi_{0}\rangle=\frac{1}{\sqrt5}\begin{pmatrix} 2\\-i  \end{pmatrix} , |\psi_{f}\rangle=\frac{1}{\sqrt6}\begin{pmatrix} 1\\2+i  \end{pmatrix}$ , then  $\vec{r_{0}}=(0,-\frac{4}{5},\frac{3}{5}) , \vec{r_{f}}=(\frac{2}{3},\frac{1}{3},-\frac{2}{3})$ then $\hat{n}=(0.816,-0.571,-0.0816)$ and by $\alpha=\pi , t_{f}=10^{-12}$sec, we obtain : $\lambda=1.234 , V_{0}=-0.00715 , V_{g0}=0.00093 , \omega_{c}=703035393816$. I mention here that for $\omega_{c}<<(\omega_{z}-\omega_{x})$ or  $\omega_{c}<<(\omega_{z}+\omega_{x})$ , all sine and cosine terms cancel each other and the qubit state is not controllable by microwave signal. In figure\ref{charge_control} the initial state, final state and two trajectories of states on Bloch sphere are depicted. The controller is designed for approximated charge qubit, then the result of control action on approximated system along blue trajectory reach to final desired state, but really the main charge qubit system has nonlinearity which exist in reality. The result of control signal on this actual system along red trajectory reach to state $|\psi_{P}\rangle$ which completely has large error to desired final state. At time $t_{f}$ all control signals are zero and in charge qubit, without control, the quantum state remains at the final state.

with same strategy for phase qubit we can design controller for approximated system, for this consider  $\hat{H}_{c}=-\frac{\hslash}{2e}\phi_{zpf}I_{g}\sigma_{x}$ that as $I_{g}(t)=I_{ac}+I_{dc}=I_{0}s(t)\mathrm{sin}(\omega_{c}t+\lambda)+I_{g0}$, the rf-controlled Hamiltonian as $\hat{H}_{rf}=U^{\dag}_{0}kI_{0}s(t)\mathrm{sin}(\omega_{c}t+\lambda)\sigma_{x}U_{0}$ that here $k=-\frac{\hslash}{2e}\phi_{zpf}$ and $U_{0}=e^{-\frac{it}{\hslash}(-\frac{1}{2}E_{c}\sigma_{z}+\frac{1}{2}E_{J}\sigma_{x})}=e^{\frac{i}{2}\omega_{z}t\sigma_{z}-\frac{i}{2}\omega_{x}t\sigma_{x}}$, with similar approach to charge qubit, multiplication $\hat{H}_{rf}$ has sine and cosine terms with frequencies $\omega_{c}-(\omega_{x}-\omega_{z})$ , $\omega_{c}+\omega_{x}+\omega_{z}$ , $\omega_{c}+\omega_{x}-\omega_{z}$ , $\omega_{c}-\omega_{x}-\omega_{z}$ and since for phase qubit $\omega_{x}>>\omega_{z}$ by rotating wave approximation, only first term is hold and other three fast rotating terms and constant offset can be ignored and setting $\delta\omega=\omega_{c}-(\omega_{x}-\omega_{z})=0$, then the rf-controlled Hamiltonian is:
\begin{align}
\begin{aligned}
\label{equ:1}
\hat{H}_{rf}=\frac{_1}{^{16}}kI_{0}s(t)(Q\sigma_{x}-4I\sigma_{y}-4Q\sigma_{z})
\end{aligned}
\end{align}
By adding constant dc current pulse to change the total control propagator operator to:
 \begin{align}
\begin{aligned}
\label{equ:1}
  U_{c}(t)=e^{\frac{-ik}{\hslash}(\frac{I_{0}}{16}\gamma(t)Q\sigma_{x}-\frac{I_{0}}{4}\gamma(t)I\sigma_{y}+(-\frac{I_{0}}{4}\gamma(t)I+I_{g0}t)\sigma_{z})}
\end{aligned}
\end{align}  

comparing to rotation propagator parameters, it obtains: $\omega_{q}n_{x}=\frac{1}{8\hslash}kI_{0}Q , \omega_{q}n_{y}=-\frac{1}{2\hslash}kI_{0}I , \omega_{q}n_{z}=-\frac{1}{2\hslash}kI_{0}Q+\frac{2}{\hslash}kI_{g0}$ then the whole signal parameters are: $\lambda=tan^{-1}(-\frac{4n_{x}}{n_{y}}) , I_{0}=\frac{8\hslash\omega_{q} n_{x}}{kQ} , I_{g0}=\frac{\hslash\omega_{q}}{2k}(n_{z}+4n_{x})$.
For example, let set the initial and final states as $|\psi_{0}\rangle=\frac{1}{\sqrt{22}}\begin{pmatrix} -2\\3-3i  \end{pmatrix} , |\psi_{f}\rangle=\frac{1}{3}\begin{pmatrix} 2\\2-i  \end{pmatrix}$ , then  $\vec{r_{0}}=(0.545,0.545,-0.636) , \vec{r_{f}}=(\frac{8}{9},-\frac{4}{9},-\frac{1}{9})$ then $\hat{n}=(0.414,0.122, -0.902)$ and by $\alpha=\pi , t_{f}=10^{-12}$sec, we obtain : $\lambda=-1.189 , I_{0}=-0.00011 , I_{g0}=-2.926\times10^{-5} , \omega_{c}=309697312116$. Figure\ref{phase_control} shows the result of simulations for this selection of initial and final states. 

The flux qubit approximate Hamiltonian is completely similar to phase qubit approximate Hamiltonian with this difference that only control derive is flux (phase), hence by considering $\hat{H}_{c}=-E_{L}\phi_{zpf}\phi_{e}\sigma_{x}$ that as $\phi_{e}(t)=\phi_{ac}+\phi_{dc}=\phi_{0}s(t)\mathrm{sin}(\omega_{c}t+\lambda)+\phi_{dc}$, the rf-controlled Hamiltonian as  $\hat{H}_{rf}=U^{\dag}_{0}k\phi_{0}s(t)\mathrm{sin}(\omega_{c}t+\lambda)\sigma_{x}U_{0}$  in which $U_{0}=e^{-\frac{it}{\hslash}(\frac{1}{2}E_{c}\sigma_{z}+\frac{1}{2}E_{J}\sigma_{x})}=e^{-\frac{i}{2}\omega_{z}t\sigma_{z}-\frac{i}{2}\omega_{x}t\sigma_{x}}$ and $k=-E_{L}\phi_{zpf}$, by simplification of sine, cosine products, we will have the frequencies  $\omega_{c}-(\omega_{x}-\omega_{z})$ , $\omega_{c}+\omega_{x}+\omega_{z}$ , $\omega_{c}+\omega_{x}-\omega_{z}$ , $\omega_{c}-\omega_{x}-\omega_{z}$ and since for flux qubit $\omega_{x}>\omega_{z}$ by rotating wave approximation, only first term is hold and other three fast rotating terms and constant offset can be ignored and setting $\delta\omega=\omega_{c}-(\omega_{x}-\omega_{z})=0$, then the rf-controlled Hamiltonian is:
\begin{align}
\begin{aligned}
\label{equ:1}
\hat{H}_{rf}=\frac{_1}{^{16}}k\phi_{0}s(t)(Q\sigma_{x}-4I\sigma_{y}-4Q\sigma_{z})
\end{aligned}
\end{align}
By adding constant dc flux pulse to change the total control propagator operator to:
 \begin{align}
\begin{aligned}
\label{equ:1}
  U_{c}(t)=e^{\frac{-ik}{\hslash}(\frac{\phi_{0}}{16}\gamma(t)Q\sigma_{x}-\frac{\phi_{0}}{4}\gamma(t)I\sigma_{y}+(-\frac{\phi_{0}}{4}\gamma(t)I+\phi_{dc}t)\sigma_{z})}
\end{aligned}
\end{align}  

the whole signal parameters are: $\lambda=tan^{-1}(-\frac{4n_{x}}{n_{y}}) , \phi_{0}=\frac{8\hslash\omega_{q} n_{x}}{kQ} , \phi_{dc}=\frac{\hslash\omega_{q}}{2k}(n_{z}+4n_{x})$. For example, let set the initial and final states as $|\psi_{0}\rangle=\frac{1}{\sqrt2}\begin{pmatrix} -1\\i  \end{pmatrix} , |\psi_{f}\rangle=\frac{1}{\sqrt{69}}\begin{pmatrix} 2\\1+8i  \end{pmatrix}$ , then  $\vec{r_{0}}=(0,-1,0) , \vec{r_{f}}=(0.058,0.463,-0.884)$ then $\hat{n}=(0.056,-0.517,-0.853)$ and by $\alpha=\pi , t_{f}=10^{-12}$sec, we obtain : $\lambda=1.0122 , \phi_{0}=-19.162 , \phi_{dc}=-3.554 , \omega_{c}=408312678393$. 

\begin{figure}[ht]
	\vspace{-0.1cm}
	\includegraphics[width=0.7 \linewidth]{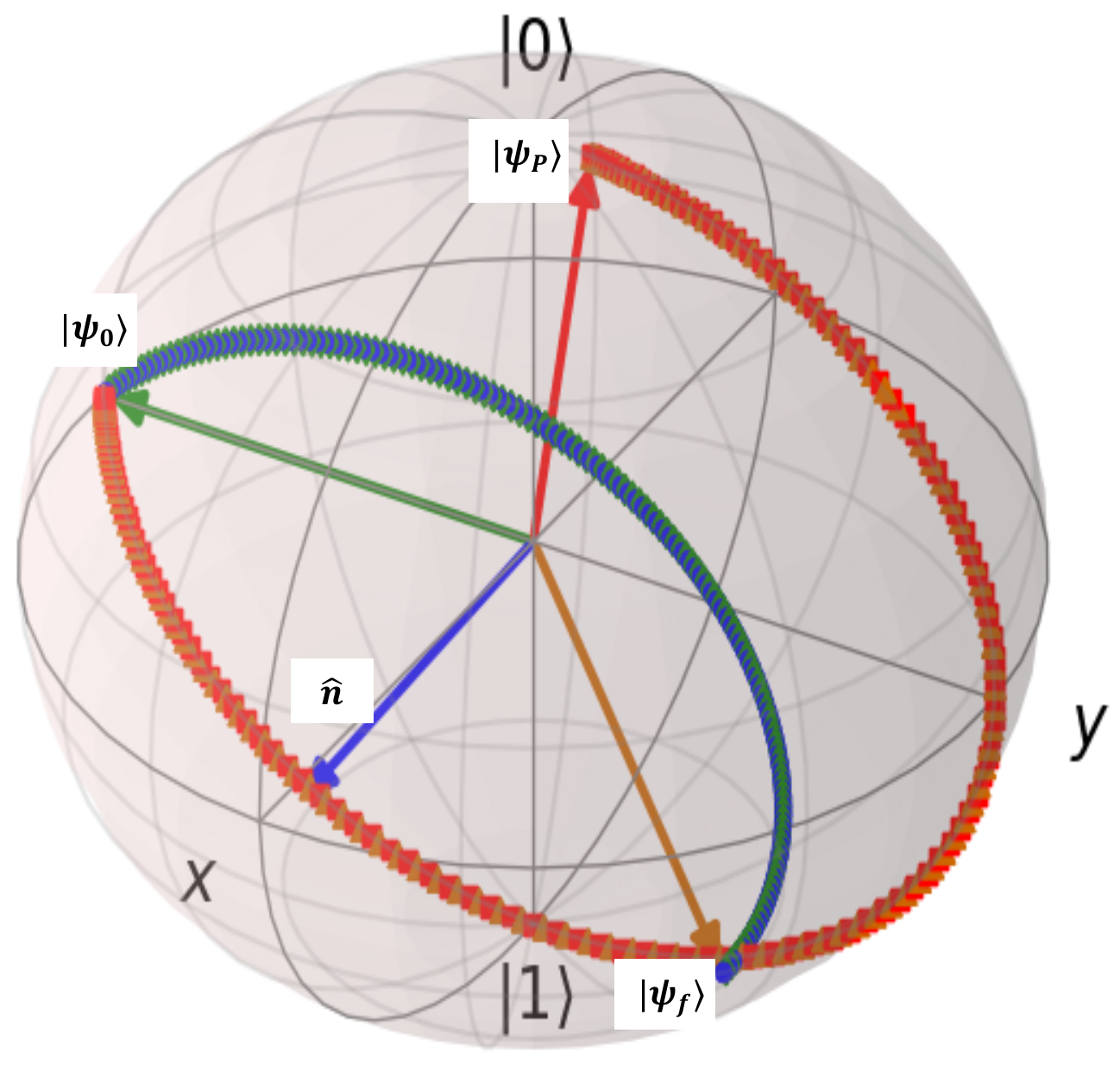}
	\vspace{-0.4cm}
	\caption{Quantum state control of superconductor flux qubit: $|\psi_{0}\rangle$ is initial state $|\psi_{f}\rangle$ is final desired state. Controlled desired trajectory is depicted in Blue. Exerting the designed controller to main system move trough red trajectory and reach to wrong state $|\psi_{P}\rangle$.} 
	\label{flux_control}
\end{figure}

As it is illustrated in figure\ref{flux_control}, the controller rotate initial state several times and finally stop to wrong position with respect to desired final state.

\section{L-C-JJ Qubit Control}
Let compute the master equation for a charge-phase-flux qubit (parallel L-JJ-C circuit) for analysis and control purpose, the general controlled Hamiltonian is:
\begin{equation}
	H=E_{c}(\hat{n}-n_{g})^{2}-E_{J}\mathrm{cos}(\hat{\phi})+\frac{1}{2}E_{L} (\hat{\phi}-\phi_{e})^2-\frac{\hslash}{2e}I_{g}\hat{\phi}
\end{equation}
let in first formulation get all of control signals in this charge-phase-flux qubit, by using $n_{g}=\frac{1}{2}C_{g}/eV(t), \phi_{e}=\phi(t),I_{g}=I(t)$, and with $\hat{\phi}=\phi_{zpf}\sigma_{x},\hat{n}=n_{zpf}\sigma_{y}$ , after computing the master equation, the following bi-linear differential equations arise:
\begin{equation}
	\begin{array}{lr}
		\label{9}
		\dot{x}(t)=-\frac{n_{zpf}}{2e\hslash}E_{c}C_{g}V(t)z(t)\\
		\dot{y}(t)=\phi_{zpf}(\frac{2}{\hslash}E_{L}\phi(t)+\frac{1}{e}I(t))z(t)\\
		\dot{z}(t)=\frac{n_{zpf}}{2e\hslash}E_{c}C_{g}V(t)x(t)
		-\phi_{zpf}(\frac{2}{\hslash}E_{L}\phi(t)+\frac{1}{e}I(t))y(t)
	\end{array}
\end{equation}

It found by this equation that in absence of external voltage, current or flux, the system for all time remain to initial state. Also, it is clear that for full control of states, it only needs control voltage $V(t)$ and one of the flux signal or current signal. let us set the external control flux signal be zero and controlling the system based on $V(t),I(t)$. Equilibrium state is a state at which without control signals, system relax. For this system, all points are an equilibrium state.
A control method for general quantum nonlinear systems is the  Lyapunov function\cite{sharifi2011lyapunov,hou2012optimal,wang2010analysis}. In Lyapunov method, must find a positive definite scalar function and  control signals must be selected such that the time derivative of Lyapunov function be negative definite, in this situation, it grantees that the initial state stabilizes toward a desired state. Let $(x_{f},y_{f},z_{f})$ be the final state and $\vec{e}=(x_{t}-x_{f},y_{t}-y_{f},z_{t}-z_{f})$ be the error between state and final state, define the Lyapunov function as Euclidean norm of error $\gamma(\vec{e})=\frac{1}{2}|\vec{e}|^2$, hence: $\dot{\gamma}(\vec{e})=\dot{\vec{e}} \vec{e}^{T}=\dot{x}_{t}(x_{t}-x_{f})+\dot{y}_{t}(y_{t}-y_{f})+\dot{z}_{t}(z_{t}-z_{f})$, then by this and equation\ref{9} and setting $\phi_{t}=0$, it obtains: $\dot{\gamma}(\vec{e})=-\frac{1}{2e\hslash}n_{zpf}E_{c}C_{g}V(t)(x_{t}z_{f}-x_{f}z_{t})-\frac{\phi_{zpf}}{e}I(t)(y_{f}z_{t}-y_{t}z_{f})$, then by selecting the linear control voltage as $V(t)=\frac{2\alpha e\hslash}{E_{c}n_{zpf}C_{g}}(x_{t}z_{f}-x_{f}z_{t})$ and linear control current as $I(t)=\frac{\beta e}{\phi_{zpf}}(y_{f}z_{t}-y_{t}z_{f})$  for $\alpha , \beta>0$ then $\dot{\gamma}(\vec{e})=-\alpha(x_{t}z_{f}-x_{f}z_{t})^2-\beta(y_{f}z_{t}-y_{t}z_{f})^2<0$ and the L-C-JJ qubit state will converge to final state and stabilizes. 
\begin{figure}[ht]
	\vspace{-0.1cm}
	\includegraphics[width=0.9 \linewidth]{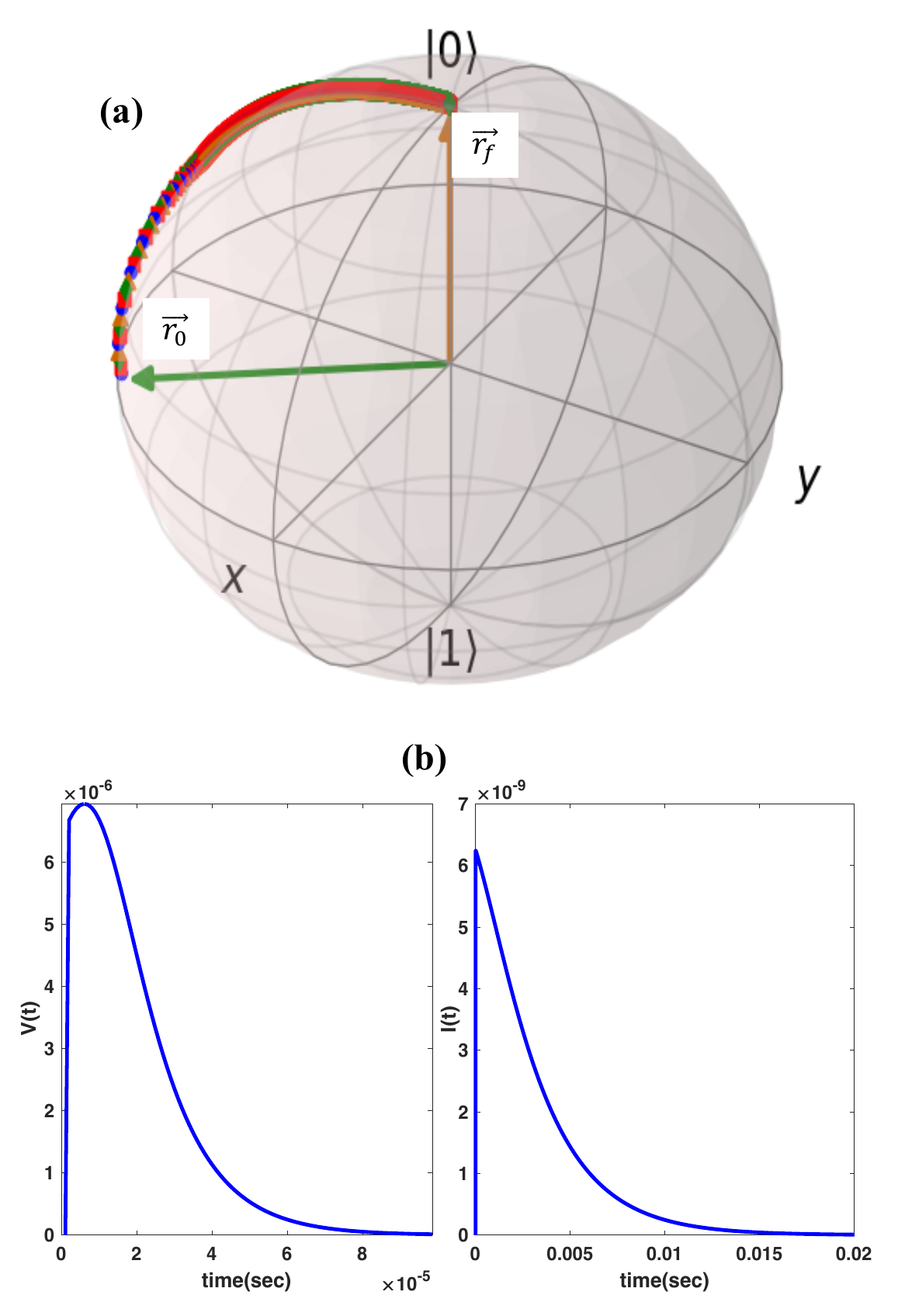}
	\vspace{-0.4cm}
	\caption{Trajectory of state control with Lyapunov control on Bloch sphere (a) and control signals (b)} 
	\label{Lyapunov}
\end{figure}
Simulations for initial state $\vec{r}_{0}=(4/9,-8/9,-1/9)$and final desired state $\vec{r}_{f}=(0,0,1)$ is depicted in figure\ref{Lyapunov}. This simulation has 20000 samples and $dt=1\mu$sec, $\alpha=\frac{1}{5}\beta=1\times10^{10}$. As it figure out from figure\ref{Lyapunov}-(b), the voltage control is of order $\mu$v and the current control has  order of nA. I mention that Lyapunov control guarantee to reach to desired state, however it does not consider the control performance criteria, hence  for this aid other control methods such as quantum optimal control or quantum reinforcement learning control can be developed. \\

\appendix

\bibliography{References}

\begin{thebibliography}{26}%
\makeatletter
\providecommand \@ifxundefined [1]{%
 \@ifx{#1\undefined}
}%
\providecommand \@ifnum [1]{%
 \ifnum #1\expandafter \@firstoftwo
 \else \expandafter \@secondoftwo
 \fi
}%
\providecommand \@ifx [1]{%
 \ifx #1\expandafter \@firstoftwo
 \else \expandafter \@secondoftwo
 \fi
}%
\providecommand \natexlab [1]{#1}%
\providecommand \enquote  [1]{``#1''}%
\providecommand \bibnamefont  [1]{#1}%
\providecommand \bibfnamefont [1]{#1}%
\providecommand \citenamefont [1]{#1}%
\providecommand \href@noop [0]{\@secondoftwo}%
\providecommand \href [0]{\begingroup \@sanitize@url \@href}%
\providecommand \@href[1]{\@@startlink{#1}\@@href}%
\providecommand \@@href[1]{\endgroup#1\@@endlink}%
\providecommand \@sanitize@url [0]{\catcode `\\12\catcode `\$12\catcode
  `\&12\catcode `\#12\catcode `\^12\catcode `\_12\catcode `\%12\relax}%
\providecommand \@@startlink[1]{}%
\providecommand \@@endlink[0]{}%
\providecommand \url  [0]{\begingroup\@sanitize@url \@url }%
\providecommand \@url [1]{\endgroup\@href {#1}{\urlprefix }}%
\providecommand \urlprefix  [0]{URL }%
\providecommand \Eprint [0]{\href }%
\providecommand \doibase [0]{https://doi.org/}%
\providecommand \selectlanguage [0]{\@gobble}%
\providecommand \bibinfo  [0]{\@secondoftwo}%
\providecommand \bibfield  [0]{\@secondoftwo}%
\providecommand \translation [1]{[#1]}%
\providecommand \BibitemOpen [0]{}%
\providecommand \bibitemStop [0]{}%
\providecommand \bibitemNoStop [0]{.\EOS\space}%
\providecommand \EOS [0]{\spacefactor3000\relax}%
\providecommand \BibitemShut  [1]{\csname bibitem#1\endcsname}%
\let\auto@bib@innerbib\@empty
\bibitem [{\citenamefont {Cirac}\ and\ \citenamefont
  {Zoller}(1995)}]{cirac1995quantum}%
  \BibitemOpen
  \bibfield  {author} {\bibinfo {author} {\bibfnamefont {J.~I.}\ \bibnamefont
  {Cirac}}\ and\ \bibinfo {author} {\bibfnamefont {P.}~\bibnamefont {Zoller}},\
  }\bibfield  {title} {\bibinfo {title} {Quantum computations with cold trapped
  ions},\ }\href@noop {} {\bibfield  {journal} {\bibinfo  {journal} {Physical
  review letters}\ }\textbf {\bibinfo {volume} {74}},\ \bibinfo {pages} {4091}
  (\bibinfo {year} {1995})}\BibitemShut {NoStop}%
\bibitem [{\citenamefont {Blatt}\ and\ \citenamefont
  {Wineland}(2008)}]{blatt2008entangled}%
  \BibitemOpen
  \bibfield  {author} {\bibinfo {author} {\bibfnamefont {R.}~\bibnamefont
  {Blatt}}\ and\ \bibinfo {author} {\bibfnamefont {D.}~\bibnamefont
  {Wineland}},\ }\bibfield  {title} {\bibinfo {title} {Entangled states of
  trapped atomic ions},\ }\href@noop {} {\bibfield  {journal} {\bibinfo
  {journal} {Nature}\ }\textbf {\bibinfo {volume} {453}},\ \bibinfo {pages}
  {1008} (\bibinfo {year} {2008})}\BibitemShut {NoStop}%
\bibitem [{\citenamefont {Pla}\ \emph {et~al.}(2013)\citenamefont {Pla},
  \citenamefont {Tan}, \citenamefont {Dehollain}, \citenamefont {Lim},
  \citenamefont {Morton}, \citenamefont {Zwanenburg}, \citenamefont {Jamieson},
  \citenamefont {Dzurak},\ and\ \citenamefont {Morello}}]{pla2013high}%
  \BibitemOpen
  \bibfield  {author} {\bibinfo {author} {\bibfnamefont {J.~J.}\ \bibnamefont
  {Pla}}, \bibinfo {author} {\bibfnamefont {K.~Y.}\ \bibnamefont {Tan}},
  \bibinfo {author} {\bibfnamefont {J.~P.}\ \bibnamefont {Dehollain}}, \bibinfo
  {author} {\bibfnamefont {W.~H.}\ \bibnamefont {Lim}}, \bibinfo {author}
  {\bibfnamefont {J.~J.}\ \bibnamefont {Morton}}, \bibinfo {author}
  {\bibfnamefont {F.~A.}\ \bibnamefont {Zwanenburg}}, \bibinfo {author}
  {\bibfnamefont {D.~N.}\ \bibnamefont {Jamieson}}, \bibinfo {author}
  {\bibfnamefont {A.~S.}\ \bibnamefont {Dzurak}},\ and\ \bibinfo {author}
  {\bibfnamefont {A.}~\bibnamefont {Morello}},\ }\bibfield  {title} {\bibinfo
  {title} {High-fidelity readout and control of a nuclear spin qubit in
  silicon},\ }\href@noop {} {\bibfield  {journal} {\bibinfo  {journal}
  {Nature}\ }\textbf {\bibinfo {volume} {496}},\ \bibinfo {pages} {334}
  (\bibinfo {year} {2013})}\BibitemShut {NoStop}%
\bibitem [{\citenamefont {Sharifi}\ and\ \citenamefont
  {Momeni}(2011)}]{sharifi2011lyapunov}%
  \BibitemOpen
  \bibfield  {author} {\bibinfo {author} {\bibfnamefont {J.}~\bibnamefont
  {Sharifi}}\ and\ \bibinfo {author} {\bibfnamefont {H.}~\bibnamefont
  {Momeni}},\ }\bibfield  {title} {\bibinfo {title} {Lyapunov control of
  squeezed noise quantum trajectory},\ }\href@noop {} {\bibfield  {journal}
  {\bibinfo  {journal} {Physics Letters A}\ }\textbf {\bibinfo {volume}
  {375}},\ \bibinfo {pages} {522} (\bibinfo {year} {2011})}\BibitemShut
  {NoStop}%
\bibitem [{\citenamefont {Pla}\ \emph {et~al.}(2012)\citenamefont {Pla},
  \citenamefont {Tan}, \citenamefont {Dehollain}, \citenamefont {Lim},
  \citenamefont {Morton}, \citenamefont {Jamieson}, \citenamefont {Dzurak},\
  and\ \citenamefont {Morello}}]{pla2012single}%
  \BibitemOpen
  \bibfield  {author} {\bibinfo {author} {\bibfnamefont {J.~J.}\ \bibnamefont
  {Pla}}, \bibinfo {author} {\bibfnamefont {K.~Y.}\ \bibnamefont {Tan}},
  \bibinfo {author} {\bibfnamefont {J.~P.}\ \bibnamefont {Dehollain}}, \bibinfo
  {author} {\bibfnamefont {W.~H.}\ \bibnamefont {Lim}}, \bibinfo {author}
  {\bibfnamefont {J.~J.}\ \bibnamefont {Morton}}, \bibinfo {author}
  {\bibfnamefont {D.~N.}\ \bibnamefont {Jamieson}}, \bibinfo {author}
  {\bibfnamefont {A.~S.}\ \bibnamefont {Dzurak}},\ and\ \bibinfo {author}
  {\bibfnamefont {A.}~\bibnamefont {Morello}},\ }\bibfield  {title} {\bibinfo
  {title} {A single-atom electron spin qubit in silicon},\ }\href@noop {}
  {\bibfield  {journal} {\bibinfo  {journal} {Nature}\ }\textbf {\bibinfo
  {volume} {489}},\ \bibinfo {pages} {541} (\bibinfo {year}
  {2012})}\BibitemShut {NoStop}%
\bibitem [{\citenamefont {Bonadeo}\ \emph {et~al.}(1998)\citenamefont
  {Bonadeo}, \citenamefont {Erland}, \citenamefont {Gammon}, \citenamefont
  {Park}, \citenamefont {Katzer},\ and\ \citenamefont
  {Steel}}]{bonadeo1998coherent}%
  \BibitemOpen
  \bibfield  {author} {\bibinfo {author} {\bibfnamefont {N.~H.}\ \bibnamefont
  {Bonadeo}}, \bibinfo {author} {\bibfnamefont {J.}~\bibnamefont {Erland}},
  \bibinfo {author} {\bibfnamefont {D.}~\bibnamefont {Gammon}}, \bibinfo
  {author} {\bibfnamefont {D.}~\bibnamefont {Park}}, \bibinfo {author}
  {\bibfnamefont {D.}~\bibnamefont {Katzer}},\ and\ \bibinfo {author}
  {\bibfnamefont {D.}~\bibnamefont {Steel}},\ }\bibfield  {title} {\bibinfo
  {title} {Coherent optical control of the quantum state of a single quantum
  dot},\ }\href@noop {} {\bibfield  {journal} {\bibinfo  {journal} {Science}\
  }\textbf {\bibinfo {volume} {282}},\ \bibinfo {pages} {1473} (\bibinfo {year}
  {1998})}\BibitemShut {NoStop}%
\bibitem [{\citenamefont {Chen}\ \emph {et~al.}(2004)\citenamefont {Chen},
  \citenamefont {Piermarocchi}, \citenamefont {Sham}, \citenamefont {Gammon},\
  and\ \citenamefont {Steel}}]{chen2004theory}%
  \BibitemOpen
  \bibfield  {author} {\bibinfo {author} {\bibfnamefont {P.}~\bibnamefont
  {Chen}}, \bibinfo {author} {\bibfnamefont {C.}~\bibnamefont {Piermarocchi}},
  \bibinfo {author} {\bibfnamefont {L.}~\bibnamefont {Sham}}, \bibinfo {author}
  {\bibfnamefont {D.}~\bibnamefont {Gammon}},\ and\ \bibinfo {author}
  {\bibfnamefont {D.}~\bibnamefont {Steel}},\ }\bibfield  {title} {\bibinfo
  {title} {Theory of quantum optical control of a single spin in a quantum
  dot},\ }\href@noop {} {\bibfield  {journal} {\bibinfo  {journal} {Physical
  Review B}\ }\textbf {\bibinfo {volume} {69}},\ \bibinfo {pages} {075320}
  (\bibinfo {year} {2004})}\BibitemShut {NoStop}%
\bibitem [{\citenamefont {Greilich}\ \emph {et~al.}(2006)\citenamefont
  {Greilich}, \citenamefont {Oulton}, \citenamefont {Zhukov}, \citenamefont
  {Yugova}, \citenamefont {Yakovlev}, \citenamefont {Bayer}, \citenamefont
  {Shabaev}, \citenamefont {Efros}, \citenamefont {Merkulov}, \citenamefont
  {Stavarache} \emph {et~al.}}]{greilich2006optical}%
  \BibitemOpen
  \bibfield  {author} {\bibinfo {author} {\bibfnamefont {A.}~\bibnamefont
  {Greilich}}, \bibinfo {author} {\bibfnamefont {R.}~\bibnamefont {Oulton}},
  \bibinfo {author} {\bibfnamefont {E.}~\bibnamefont {Zhukov}}, \bibinfo
  {author} {\bibfnamefont {I.}~\bibnamefont {Yugova}}, \bibinfo {author}
  {\bibfnamefont {D.}~\bibnamefont {Yakovlev}}, \bibinfo {author}
  {\bibfnamefont {M.}~\bibnamefont {Bayer}}, \bibinfo {author} {\bibfnamefont
  {A.}~\bibnamefont {Shabaev}}, \bibinfo {author} {\bibfnamefont {A.~L.}\
  \bibnamefont {Efros}}, \bibinfo {author} {\bibfnamefont {I.}~\bibnamefont
  {Merkulov}}, \bibinfo {author} {\bibfnamefont {V.}~\bibnamefont
  {Stavarache}}, \emph {et~al.},\ }\bibfield  {title} {\bibinfo {title}
  {Optical control of spin coherence in singly charged (in, ga) as/gaas quantum
  dots},\ }\href@noop {} {\bibfield  {journal} {\bibinfo  {journal} {Physical
  review letters}\ }\textbf {\bibinfo {volume} {96}},\ \bibinfo {pages}
  {227401} (\bibinfo {year} {2006})}\BibitemShut {NoStop}%
\bibitem [{\citenamefont {Wallraff}\ \emph {et~al.}(2005)\citenamefont
  {Wallraff}, \citenamefont {Schuster}, \citenamefont {Blais}, \citenamefont
  {Frunzio}, \citenamefont {Majer}, \citenamefont {Devoret}, \citenamefont
  {Girvin},\ and\ \citenamefont {Schoelkopf}}]{wallraff2005approaching}%
  \BibitemOpen
  \bibfield  {author} {\bibinfo {author} {\bibfnamefont {A.}~\bibnamefont
  {Wallraff}}, \bibinfo {author} {\bibfnamefont {D.}~\bibnamefont {Schuster}},
  \bibinfo {author} {\bibfnamefont {A.}~\bibnamefont {Blais}}, \bibinfo
  {author} {\bibfnamefont {L.}~\bibnamefont {Frunzio}}, \bibinfo {author}
  {\bibfnamefont {J.}~\bibnamefont {Majer}}, \bibinfo {author} {\bibfnamefont
  {M.}~\bibnamefont {Devoret}}, \bibinfo {author} {\bibfnamefont
  {S.}~\bibnamefont {Girvin}},\ and\ \bibinfo {author} {\bibfnamefont
  {R.}~\bibnamefont {Schoelkopf}},\ }\bibfield  {title} {\bibinfo {title}
  {Approaching unit visibility for control of a superconducting qubit with
  dispersive readout},\ }\href@noop {} {\bibfield  {journal} {\bibinfo
  {journal} {Physical review letters}\ }\textbf {\bibinfo {volume} {95}},\
  \bibinfo {pages} {060501} (\bibinfo {year} {2005})}\BibitemShut {NoStop}%
\bibitem [{\citenamefont {Campagne-Ibarcq}\ \emph {et~al.}(2013)\citenamefont
  {Campagne-Ibarcq}, \citenamefont {Flurin}, \citenamefont {Roch},
  \citenamefont {Darson}, \citenamefont {Morfin}, \citenamefont {Mirrahimi},
  \citenamefont {Devoret}, \citenamefont {Mallet},\ and\ \citenamefont
  {Huard}}]{campagne2013persistent}%
  \BibitemOpen
  \bibfield  {author} {\bibinfo {author} {\bibfnamefont {P.}~\bibnamefont
  {Campagne-Ibarcq}}, \bibinfo {author} {\bibfnamefont {E.}~\bibnamefont
  {Flurin}}, \bibinfo {author} {\bibfnamefont {N.}~\bibnamefont {Roch}},
  \bibinfo {author} {\bibfnamefont {D.}~\bibnamefont {Darson}}, \bibinfo
  {author} {\bibfnamefont {P.}~\bibnamefont {Morfin}}, \bibinfo {author}
  {\bibfnamefont {M.}~\bibnamefont {Mirrahimi}}, \bibinfo {author}
  {\bibfnamefont {M.~H.}\ \bibnamefont {Devoret}}, \bibinfo {author}
  {\bibfnamefont {F.}~\bibnamefont {Mallet}},\ and\ \bibinfo {author}
  {\bibfnamefont {B.}~\bibnamefont {Huard}},\ }\bibfield  {title} {\bibinfo
  {title} {Persistent control of a superconducting qubit by stroboscopic
  measurement feedback},\ }\href@noop {} {\bibfield  {journal} {\bibinfo
  {journal} {Physical Review X}\ }\textbf {\bibinfo {volume} {3}},\ \bibinfo
  {pages} {021008} (\bibinfo {year} {2013})}\BibitemShut {NoStop}%
\bibitem [{\citenamefont {Leonard~Jr}\ \emph {et~al.}(2019)\citenamefont
  {Leonard~Jr}, \citenamefont {Beck}, \citenamefont {Nelson}, \citenamefont
  {Christensen}, \citenamefont {Thorbeck}, \citenamefont {Howington},
  \citenamefont {Opremcak}, \citenamefont {Pechenezhskiy}, \citenamefont
  {Dodge}, \citenamefont {Dupuis} \emph {et~al.}}]{leonard2019digital}%
  \BibitemOpen
  \bibfield  {author} {\bibinfo {author} {\bibfnamefont {E.}~\bibnamefont
  {Leonard~Jr}}, \bibinfo {author} {\bibfnamefont {M.~A.}\ \bibnamefont
  {Beck}}, \bibinfo {author} {\bibfnamefont {J.}~\bibnamefont {Nelson}},
  \bibinfo {author} {\bibfnamefont {B.~G.}\ \bibnamefont {Christensen}},
  \bibinfo {author} {\bibfnamefont {T.}~\bibnamefont {Thorbeck}}, \bibinfo
  {author} {\bibfnamefont {C.}~\bibnamefont {Howington}}, \bibinfo {author}
  {\bibfnamefont {A.}~\bibnamefont {Opremcak}}, \bibinfo {author}
  {\bibfnamefont {I.~V.}\ \bibnamefont {Pechenezhskiy}}, \bibinfo {author}
  {\bibfnamefont {K.}~\bibnamefont {Dodge}}, \bibinfo {author} {\bibfnamefont
  {N.~P.}\ \bibnamefont {Dupuis}}, \emph {et~al.},\ }\bibfield  {title}
  {\bibinfo {title} {Digital coherent control of a superconducting qubit},\
  }\href@noop {} {\bibfield  {journal} {\bibinfo  {journal} {Physical Review
  Applied}\ }\textbf {\bibinfo {volume} {11}},\ \bibinfo {pages} {014009}
  (\bibinfo {year} {2019})}\BibitemShut {NoStop}%
\bibitem [{\citenamefont {Wallraff}\ \emph {et~al.}(2004)\citenamefont
  {Wallraff}, \citenamefont {Schuster}, \citenamefont {Blais}, \citenamefont
  {Frunzio}, \citenamefont {Huang}, \citenamefont {Majer}, \citenamefont
  {Kumar}, \citenamefont {Girvin},\ and\ \citenamefont
  {Schoelkopf}}]{wallraff2004strong}%
  \BibitemOpen
  \bibfield  {author} {\bibinfo {author} {\bibfnamefont {A.}~\bibnamefont
  {Wallraff}}, \bibinfo {author} {\bibfnamefont {D.~I.}\ \bibnamefont
  {Schuster}}, \bibinfo {author} {\bibfnamefont {A.}~\bibnamefont {Blais}},
  \bibinfo {author} {\bibfnamefont {L.}~\bibnamefont {Frunzio}}, \bibinfo
  {author} {\bibfnamefont {R.-S.}\ \bibnamefont {Huang}}, \bibinfo {author}
  {\bibfnamefont {J.}~\bibnamefont {Majer}}, \bibinfo {author} {\bibfnamefont
  {S.}~\bibnamefont {Kumar}}, \bibinfo {author} {\bibfnamefont {S.~M.}\
  \bibnamefont {Girvin}},\ and\ \bibinfo {author} {\bibfnamefont {R.~J.}\
  \bibnamefont {Schoelkopf}},\ }\bibfield  {title} {\bibinfo {title} {Strong
  coupling of a single photon to a superconducting qubit using circuit quantum
  electrodynamics},\ }\href@noop {} {\bibfield  {journal} {\bibinfo  {journal}
  {Nature}\ }\textbf {\bibinfo {volume} {431}},\ \bibinfo {pages} {162}
  (\bibinfo {year} {2004})}\BibitemShut {NoStop}%
\bibitem [{\citenamefont {Houck}\ \emph {et~al.}(2012)\citenamefont {Houck},
  \citenamefont {T{\"u}reci},\ and\ \citenamefont {Koch}}]{houck2012chip}%
  \BibitemOpen
  \bibfield  {author} {\bibinfo {author} {\bibfnamefont {A.~A.}\ \bibnamefont
  {Houck}}, \bibinfo {author} {\bibfnamefont {H.~E.}\ \bibnamefont
  {T{\"u}reci}},\ and\ \bibinfo {author} {\bibfnamefont {J.}~\bibnamefont
  {Koch}},\ }\bibfield  {title} {\bibinfo {title} {On-chip quantum simulation
  with superconducting circuits},\ }\href@noop {} {\bibfield  {journal}
  {\bibinfo  {journal} {Nature Physics}\ }\textbf {\bibinfo {volume} {8}},\
  \bibinfo {pages} {292} (\bibinfo {year} {2012})}\BibitemShut {NoStop}%
\bibitem [{\citenamefont {Fedorov}\ \emph {et~al.}(2012)\citenamefont
  {Fedorov}, \citenamefont {Steffen}, \citenamefont {Baur}, \citenamefont
  {da~Silva},\ and\ \citenamefont {Wallraff}}]{fedorov2012implementation}%
  \BibitemOpen
  \bibfield  {author} {\bibinfo {author} {\bibfnamefont {A.}~\bibnamefont
  {Fedorov}}, \bibinfo {author} {\bibfnamefont {L.}~\bibnamefont {Steffen}},
  \bibinfo {author} {\bibfnamefont {M.}~\bibnamefont {Baur}}, \bibinfo {author}
  {\bibfnamefont {M.~P.}\ \bibnamefont {da~Silva}},\ and\ \bibinfo {author}
  {\bibfnamefont {A.}~\bibnamefont {Wallraff}},\ }\bibfield  {title} {\bibinfo
  {title} {Implementation of a toffoli gate with superconducting circuits},\
  }\href@noop {} {\bibfield  {journal} {\bibinfo  {journal} {Nature}\ }\textbf
  {\bibinfo {volume} {481}},\ \bibinfo {pages} {170} (\bibinfo {year}
  {2012})}\BibitemShut {NoStop}%
\bibitem [{\citenamefont {Mariantoni}\ \emph {et~al.}(2011)\citenamefont
  {Mariantoni}, \citenamefont {Wang}, \citenamefont {Yamamoto}, \citenamefont
  {Neeley}, \citenamefont {Bialczak}, \citenamefont {Chen}, \citenamefont
  {Lenander}, \citenamefont {Lucero}, \citenamefont {O’Connell},
  \citenamefont {Sank} \emph {et~al.}}]{mariantoni2011implementing}%
  \BibitemOpen
  \bibfield  {author} {\bibinfo {author} {\bibfnamefont {M.}~\bibnamefont
  {Mariantoni}}, \bibinfo {author} {\bibfnamefont {H.}~\bibnamefont {Wang}},
  \bibinfo {author} {\bibfnamefont {T.}~\bibnamefont {Yamamoto}}, \bibinfo
  {author} {\bibfnamefont {M.}~\bibnamefont {Neeley}}, \bibinfo {author}
  {\bibfnamefont {R.~C.}\ \bibnamefont {Bialczak}}, \bibinfo {author}
  {\bibfnamefont {Y.}~\bibnamefont {Chen}}, \bibinfo {author} {\bibfnamefont
  {M.}~\bibnamefont {Lenander}}, \bibinfo {author} {\bibfnamefont
  {E.}~\bibnamefont {Lucero}}, \bibinfo {author} {\bibfnamefont {A.~D.}\
  \bibnamefont {O’Connell}}, \bibinfo {author} {\bibfnamefont
  {D.}~\bibnamefont {Sank}}, \emph {et~al.},\ }\bibfield  {title} {\bibinfo
  {title} {Implementing the quantum von neumann architecture with
  superconducting circuits},\ }\href@noop {} {\bibfield  {journal} {\bibinfo
  {journal} {Science}\ }\textbf {\bibinfo {volume} {334}},\ \bibinfo {pages}
  {61} (\bibinfo {year} {2011})}\BibitemShut {NoStop}%
\bibitem [{\citenamefont {Ofek}\ \emph {et~al.}(2016)\citenamefont {Ofek},
  \citenamefont {Petrenko}, \citenamefont {Heeres}, \citenamefont {Reinhold},
  \citenamefont {Leghtas}, \citenamefont {Vlastakis}, \citenamefont {Liu},
  \citenamefont {Frunzio}, \citenamefont {Girvin}, \citenamefont {Jiang} \emph
  {et~al.}}]{ofek2016extending}%
  \BibitemOpen
  \bibfield  {author} {\bibinfo {author} {\bibfnamefont {N.}~\bibnamefont
  {Ofek}}, \bibinfo {author} {\bibfnamefont {A.}~\bibnamefont {Petrenko}},
  \bibinfo {author} {\bibfnamefont {R.}~\bibnamefont {Heeres}}, \bibinfo
  {author} {\bibfnamefont {P.}~\bibnamefont {Reinhold}}, \bibinfo {author}
  {\bibfnamefont {Z.}~\bibnamefont {Leghtas}}, \bibinfo {author} {\bibfnamefont
  {B.}~\bibnamefont {Vlastakis}}, \bibinfo {author} {\bibfnamefont
  {Y.}~\bibnamefont {Liu}}, \bibinfo {author} {\bibfnamefont {L.}~\bibnamefont
  {Frunzio}}, \bibinfo {author} {\bibfnamefont {S.}~\bibnamefont {Girvin}},
  \bibinfo {author} {\bibfnamefont {L.}~\bibnamefont {Jiang}}, \emph {et~al.},\
  }\bibfield  {title} {\bibinfo {title} {Extending the lifetime of a quantum
  bit with error correction in superconducting circuits},\ }\href@noop {}
  {\bibfield  {journal} {\bibinfo  {journal} {Nature}\ }\textbf {\bibinfo
  {volume} {536}},\ \bibinfo {pages} {441} (\bibinfo {year}
  {2016})}\BibitemShut {NoStop}%
\bibitem [{\citenamefont {Campagne-Ibarcq}\ \emph {et~al.}(2018)\citenamefont
  {Campagne-Ibarcq}, \citenamefont {Zalys-Geller}, \citenamefont {Narla},
  \citenamefont {Shankar}, \citenamefont {Reinhold}, \citenamefont {Burkhart},
  \citenamefont {Axline}, \citenamefont {Pfaff}, \citenamefont {Frunzio},
  \citenamefont {Schoelkopf} \emph {et~al.}}]{campagne2018deterministic}%
  \BibitemOpen
  \bibfield  {author} {\bibinfo {author} {\bibfnamefont {P.}~\bibnamefont
  {Campagne-Ibarcq}}, \bibinfo {author} {\bibfnamefont {E.}~\bibnamefont
  {Zalys-Geller}}, \bibinfo {author} {\bibfnamefont {A.}~\bibnamefont {Narla}},
  \bibinfo {author} {\bibfnamefont {S.}~\bibnamefont {Shankar}}, \bibinfo
  {author} {\bibfnamefont {P.}~\bibnamefont {Reinhold}}, \bibinfo {author}
  {\bibfnamefont {L.}~\bibnamefont {Burkhart}}, \bibinfo {author}
  {\bibfnamefont {C.}~\bibnamefont {Axline}}, \bibinfo {author} {\bibfnamefont
  {W.}~\bibnamefont {Pfaff}}, \bibinfo {author} {\bibfnamefont
  {L.}~\bibnamefont {Frunzio}}, \bibinfo {author} {\bibfnamefont
  {R.}~\bibnamefont {Schoelkopf}}, \emph {et~al.},\ }\bibfield  {title}
  {\bibinfo {title} {Deterministic remote entanglement of superconducting
  circuits through microwave two-photon transitions},\ }\href@noop {}
  {\bibfield  {journal} {\bibinfo  {journal} {Physical review letters}\
  }\textbf {\bibinfo {volume} {120}},\ \bibinfo {pages} {200501} (\bibinfo
  {year} {2018})}\BibitemShut {NoStop}%
\bibitem [{\citenamefont {Wendin}(2017)}]{wendin2017quantum}%
  \BibitemOpen
  \bibfield  {author} {\bibinfo {author} {\bibfnamefont {G.}~\bibnamefont
  {Wendin}},\ }\bibfield  {title} {\bibinfo {title} {Quantum information
  processing with superconducting circuits: a review},\ }\href@noop {}
  {\bibfield  {journal} {\bibinfo  {journal} {Reports on Progress in Physics}\
  }\textbf {\bibinfo {volume} {80}},\ \bibinfo {pages} {106001} (\bibinfo
  {year} {2017})}\BibitemShut {NoStop}%
\bibitem [{\citenamefont {Wendin}\ and\ \citenamefont
  {Shumeiko}(2007)}]{wendin2007quantum}%
  \BibitemOpen
  \bibfield  {author} {\bibinfo {author} {\bibfnamefont {G.}~\bibnamefont
  {Wendin}}\ and\ \bibinfo {author} {\bibfnamefont {V.}~\bibnamefont
  {Shumeiko}},\ }\bibfield  {title} {\bibinfo {title} {Quantum bits with
  josephson junctions},\ }\href@noop {} {\bibfield  {journal} {\bibinfo
  {journal} {Low Temperature Physics}\ }\textbf {\bibinfo {volume} {33}},\
  \bibinfo {pages} {724} (\bibinfo {year} {2007})}\BibitemShut {NoStop}%
\bibitem [{\citenamefont {Rodrigues}(2003)}]{rodrigues2003superconducting}%
  \BibitemOpen
  \bibfield  {author} {\bibinfo {author} {\bibfnamefont {D.~A.}\ \bibnamefont
  {Rodrigues}},\ }\emph {\bibinfo {title} {Superconducting charge qubits}},\
  \href@noop {} {Ph.D. thesis},\ \bibinfo  {school} {University of Bristol}
  (\bibinfo {year} {2003})\BibitemShut {NoStop}%
\bibitem [{\citenamefont {Saito}\ \emph {et~al.}(2006)\citenamefont {Saito},
  \citenamefont {Meno}, \citenamefont {Ueda}, \citenamefont {Tanaka},
  \citenamefont {Semba},\ and\ \citenamefont
  {Takayanagi}}]{saito2006parametric}%
  \BibitemOpen
  \bibfield  {author} {\bibinfo {author} {\bibfnamefont {S.}~\bibnamefont
  {Saito}}, \bibinfo {author} {\bibfnamefont {T.}~\bibnamefont {Meno}},
  \bibinfo {author} {\bibfnamefont {M.}~\bibnamefont {Ueda}}, \bibinfo {author}
  {\bibfnamefont {H.}~\bibnamefont {Tanaka}}, \bibinfo {author} {\bibfnamefont
  {K.}~\bibnamefont {Semba}},\ and\ \bibinfo {author} {\bibfnamefont
  {H.}~\bibnamefont {Takayanagi}},\ }\bibfield  {title} {\bibinfo {title}
  {Parametric control of a superconducting flux qubit},\ }\href@noop {}
  {\bibfield  {journal} {\bibinfo  {journal} {Physical review letters}\
  }\textbf {\bibinfo {volume} {96}},\ \bibinfo {pages} {107001} (\bibinfo
  {year} {2006})}\BibitemShut {NoStop}%
\bibitem [{\citenamefont {Kutsuzawa}\ \emph {et~al.}(2005)\citenamefont
  {Kutsuzawa}, \citenamefont {Tanaka}, \citenamefont {Saito}, \citenamefont
  {Nakano}, \citenamefont {Semba},\ and\ \citenamefont
  {Takayanagi}}]{kutsuzawa2005coherent}%
  \BibitemOpen
  \bibfield  {author} {\bibinfo {author} {\bibfnamefont {T.}~\bibnamefont
  {Kutsuzawa}}, \bibinfo {author} {\bibfnamefont {H.}~\bibnamefont {Tanaka}},
  \bibinfo {author} {\bibfnamefont {S.}~\bibnamefont {Saito}}, \bibinfo
  {author} {\bibfnamefont {H.}~\bibnamefont {Nakano}}, \bibinfo {author}
  {\bibfnamefont {K.}~\bibnamefont {Semba}},\ and\ \bibinfo {author}
  {\bibfnamefont {H.}~\bibnamefont {Takayanagi}},\ }\bibfield  {title}
  {\bibinfo {title} {Coherent control of a flux qubit by phase-shifted resonant
  microwave pulses},\ }\href@noop {} {\bibfield  {journal} {\bibinfo  {journal}
  {Applied Physics Letters}\ }\textbf {\bibinfo {volume} {87}},\ \bibinfo
  {pages} {073501} (\bibinfo {year} {2005})}\BibitemShut {NoStop}%
\bibitem [{\citenamefont {Naaman}\ \emph {et~al.}(2017)\citenamefont {Naaman},
  \citenamefont {Strong}, \citenamefont {Ferguson}, \citenamefont {Egan},
  \citenamefont {Bailey},\ and\ \citenamefont {Hinkey}}]{naaman2017josephson}%
  \BibitemOpen
  \bibfield  {author} {\bibinfo {author} {\bibfnamefont {O.}~\bibnamefont
  {Naaman}}, \bibinfo {author} {\bibfnamefont {J.}~\bibnamefont {Strong}},
  \bibinfo {author} {\bibfnamefont {D.}~\bibnamefont {Ferguson}}, \bibinfo
  {author} {\bibfnamefont {J.}~\bibnamefont {Egan}}, \bibinfo {author}
  {\bibfnamefont {N.}~\bibnamefont {Bailey}},\ and\ \bibinfo {author}
  {\bibfnamefont {R.}~\bibnamefont {Hinkey}},\ }\bibfield  {title} {\bibinfo
  {title} {Josephson junction microwave modulators for qubit control},\
  }\href@noop {} {\bibfield  {journal} {\bibinfo  {journal} {Journal of Applied
  Physics}\ }\textbf {\bibinfo {volume} {121}},\ \bibinfo {pages} {073904}
  (\bibinfo {year} {2017})}\BibitemShut {NoStop}%
\bibitem [{\citenamefont {Krantz}\ \emph {et~al.}(2019)\citenamefont {Krantz},
  \citenamefont {Kjaergaard}, \citenamefont {Yan}, \citenamefont {Orlando},
  \citenamefont {Gustavsson},\ and\ \citenamefont
  {Oliver}}]{krantz2019quantum}%
  \BibitemOpen
  \bibfield  {author} {\bibinfo {author} {\bibfnamefont {P.}~\bibnamefont
  {Krantz}}, \bibinfo {author} {\bibfnamefont {M.}~\bibnamefont {Kjaergaard}},
  \bibinfo {author} {\bibfnamefont {F.}~\bibnamefont {Yan}}, \bibinfo {author}
  {\bibfnamefont {T.~P.}\ \bibnamefont {Orlando}}, \bibinfo {author}
  {\bibfnamefont {S.}~\bibnamefont {Gustavsson}},\ and\ \bibinfo {author}
  {\bibfnamefont {W.~D.}\ \bibnamefont {Oliver}},\ }\bibfield  {title}
  {\bibinfo {title} {A quantum engineer's guide to superconducting qubits},\
  }\href@noop {} {\bibfield  {journal} {\bibinfo  {journal} {Applied Physics
  Reviews}\ }\textbf {\bibinfo {volume} {6}},\ \bibinfo {pages} {021318}
  (\bibinfo {year} {2019})}\BibitemShut {NoStop}%
\bibitem [{\citenamefont {Hou}\ \emph {et~al.}(2012)\citenamefont {Hou},
  \citenamefont {Khan}, \citenamefont {Yi}, \citenamefont {Dong},\ and\
  \citenamefont {Petersen}}]{hou2012optimal}%
  \BibitemOpen
  \bibfield  {author} {\bibinfo {author} {\bibfnamefont {S.-C.}\ \bibnamefont
  {Hou}}, \bibinfo {author} {\bibfnamefont {M.}~\bibnamefont {Khan}}, \bibinfo
  {author} {\bibfnamefont {X.}~\bibnamefont {Yi}}, \bibinfo {author}
  {\bibfnamefont {D.}~\bibnamefont {Dong}},\ and\ \bibinfo {author}
  {\bibfnamefont {I.~R.}\ \bibnamefont {Petersen}},\ }\bibfield  {title}
  {\bibinfo {title} {Optimal lyapunov-based quantum control for quantum
  systems},\ }\href@noop {} {\bibfield  {journal} {\bibinfo  {journal}
  {Physical Review A}\ }\textbf {\bibinfo {volume} {86}},\ \bibinfo {pages}
  {022321} (\bibinfo {year} {2012})}\BibitemShut {NoStop}%
\bibitem [{\citenamefont {Wang}\ and\ \citenamefont
  {Schirmer}(2010)}]{wang2010analysis}%
  \BibitemOpen
  \bibfield  {author} {\bibinfo {author} {\bibfnamefont {X.}~\bibnamefont
  {Wang}}\ and\ \bibinfo {author} {\bibfnamefont {S.~G.}\ \bibnamefont
  {Schirmer}},\ }\bibfield  {title} {\bibinfo {title} {Analysis of lyapunov
  method for control of quantum states},\ }\href@noop {} {\bibfield  {journal}
  {\bibinfo  {journal} {IEEE Transactions on Automatic control}\ }\textbf
  {\bibinfo {volume} {55}},\ \bibinfo {pages} {2259} (\bibinfo {year}
  {2010})}\BibitemShut {NoStop}%
\end{thebibliography}%


%
\end{document}